\documentclass[runningheads]{llncs}
\usepackage[T1]{fontenc}

\usepackage{graphicx}
\usepackage{hyperref}
\usepackage{amsmath}
\usepackage{amssymb}
\usepackage{booktabs}
\usepackage[table]{xcolor}
\definecolor{lightgray}{gray}{0.9}
\usepackage{siunitx}
\usepackage{multirow}
\usepackage{float}
\usepackage[misc]{ifsym}
\begin{document}
\title{Prompt Your Brain: Scaffold Prompt Tuning for Efficient Adaptation of fMRI Pre-trained Model}
\titlerunning{ScaPT}
%
\author{Zijian Dong\inst{1,2,3} \and
Yilei Wu\inst{1,2} \and
Zijiao Chen\inst{1,2} \and Yichi Zhang\inst{1,2} \and Yueming Jin\inst{3,4} \and Juan Helen Zhou\inst{1,2,3}\textsuperscript{(\Letter)}}
%
\authorrunning{Z. Dong et al.}
%
\institute{Centre for Sleep and Cognition \& Centre for Translational Magnetic Resonance Research, Yong Loo Lin School of Medicine, National University of Singapore, Singapore \and Human Potential Translational Research Program and Department of Medicine, Yong Loo Lin School of Medicine, National University of Singapore, Singapore \and Department of Electrical and Computer Engineering \& Integrative Sciences and Engineering Programme (ISEP), NUS Graduate School, National University of Singapore, Singapore \and Department of Biomedical Engineering, National University of Singapore, Singapore \\ \email{helen.zhou@nus.edu.sg}}

\maketitle              
\begin{abstract}
We introduce \textbf{Sca}ffold \textbf{P}rompt \textbf{T}uning (\textbf{ScaPT}), a novel prompt-based framework for adapting large-scale functional magnetic resonance imaging (fMRI) pre-trained models to downstream tasks, with high parameter efficiency and improved performance compared to fine-tuning and baselines for prompt tuning. The full fine-tuning updates all pre-trained parameters, which may distort the learned feature space and lead to overfitting with limited training data which is common in fMRI fields. In contrast, we design a hierarchical prompt structure that transfers the knowledge learned from high-resource tasks to low-resource ones. This structure, equipped with a Deeply-conditioned Input-Prompt (DIP) mapping module, allows for efficient adaptation by updating only 2\% of the trainable parameters. The framework enhances semantic interpretability through attention mechanisms between inputs and prompts, and it clusters prompts in the latent space in alignment with prior knowledge. Experiments on public resting state fMRI datasets reveal ScaPT outperforms fine-tuning and multitask-based prompt tuning in neurodegenerative diseases diagnosis/prognosis and personality trait prediction, even with fewer than 20 participants. It highlights ScaPT's efficiency in adapting pre-trained fMRI models to low-resource tasks.

\end{abstract}
\section{Introduction}

In the realm of neuroimaging, the emergence of large-scale, self-supervised pre-trained models/foundation models for fMRI, has demonstrated a significant promise in improving performance across a variety of downstream tasks through fully fine-tuning \cite{ortega2023brainlm,thomas2022self}. However, fine-tuning in fMRI models requires updating all the pre-trained parameters given target task training data, which is computationally intensive and time-consuming. Moreover, when confronted with a scarcity of training data, large pre-trained models are at risk of overfitting. It has also been shown that fine-tuning could distort the learned feature space \cite{sun2023fine}.

In natural language processing (NLP), the concept of soft prompt tuning presents an efficient alternative for adapting large language models (LLMs) \cite{lester2021power}. This technique involves keeping the original model frozen while only training soft prompts prepended to the input. Such an approach has been demonstrated to enhance the decoding of information represented in the human brain for language understanding using fMRI \cite{sun2023fine}. Soft prompts have also proven capable of distinguishing between different communities within brain networks \cite{bannadabhavi2023community}. Despite its efficiency, prompt tuning often results in decreased task performance compared to fine-tuning \cite{he2021towards} and fails to leverage the extensive knowledge embedded in a diverse array of high-resource tasks \cite{asai2022attempt}. Building predictive models using neuroimaging data of disease cohorts (e.g., neurodegenerative disease such as Alzheimer's disease (AD)) is typically a low-resource task due to the small sample size. Given the emerging large-scale neuroimaging data from healthy cohorts with comprehensive behavioral phenotyping, brain-behavior mapping in healthy populations is usually categorized as high-resource tasks. The knowledge gained from these high-resource tasks is invaluable and could significantly bolster the model performance in low-resource applications (e.g., disease prognosis).

Several works in NLP have designed prompt tuning strategies under multitask transfer learning framework \cite{asai2022attempt,sun2023multitask,vu2022spot}, which transfer knowledge from high-resource tasks to low-resource ones. Their source prompts are trained through multitask learning and subsequently utilized to initialize the model for a specific target task. However, they either overlook the intricate relationship between prompts and input \cite{sun2023multitask,vu2022spot}, or map an input to a task-agnostic prompt space without capturing the input's different aspects for various tasks \cite{asai2022attempt}. Furthermore, in NLP, soft prompts function as a ``black box'', lacking a semantic understanding of the information encapsulated within the embeddings. Lacking of interpretability makes it suboptimal for clinical application.

In this study, we introduce \textbf{Sca}ffold \textbf{P}rompt \textbf{T}uning (\textbf{ScaPT}), the first prompt-based adaptive framework for fMRI pre-trained model with remarkable parameter efficiency and superior performance using limited downstream training data. Our contributions are four-fold: 1) We have designed a hierarchical prompt structure that evolves soft prompts into three levels: \emph{modular prompts} at the super-domain level (treated as fundamental skills of the model) for basic fMRI knowledge embedded in the model; \emph{phenotype prompts} combining modular prompts to represent different phenotypes in the human brain; and \emph{vertex prompts} merging phenotype prompts and a newly-initialized target prompt for target tasks. Integration is guided by attention between inputs and prompts. 2) We propose a shared Deeply-conditioned Input-Prompt (DIP) mapping module for both source and target training, designed to uniquely map inputs to different prompt spaces. 3) We demonstrated that our proposed phenotype prompts form clusters, which align well with prior knowledge. Moreover, the proposed attention mechanism between inputs and prompts offers significant semantic interpretability. 4) Through experiments on two public resting state fMRI datasets, ScaPT demonstrates remarkable parameter efficiency, outperforming both fine-tuning and baseline prompt tuning methods by updating only 2\% of the trainable parameters. It achieves superior results in neurodegenerative disease diagnosis/prognosis, as well as personality trait prediction, with very limited training data.

\begin{figure}[t]
\centering
\includegraphics[width=\linewidth]{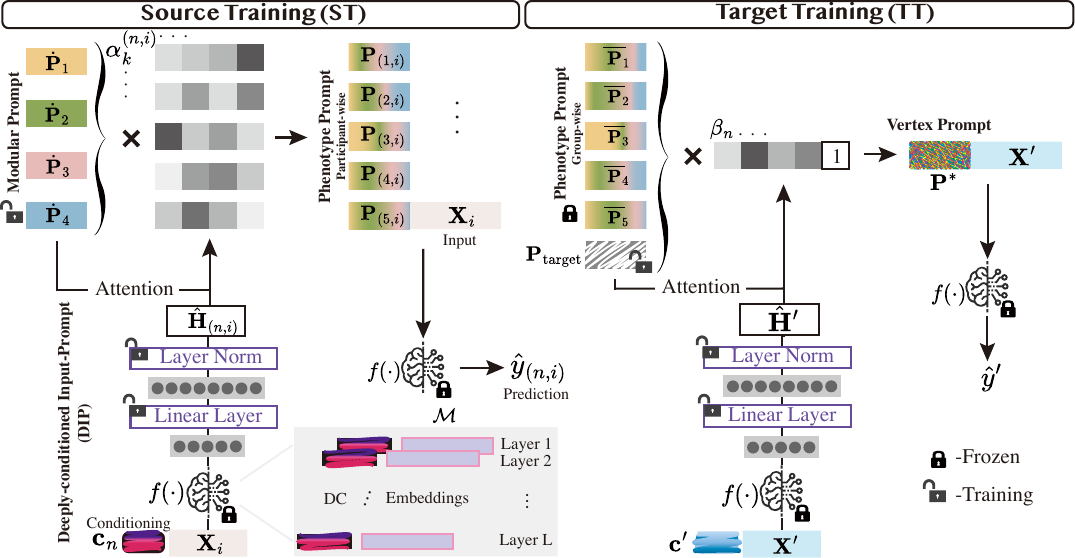}
\caption{Schematic overview of Scaffold Prompt Tuning (ScaPT) framework. ScaPT operates in two stages: \textbf{Source Training (ST)}, where it creates phenotype prompts by interpolation of modular prompts, and \textbf{Target Training (TT)}, where it blends phenotype prompts with a new target prompt for downstream tasks with fewer resources. Interpolation weights are determined by the attention between the input and prompts.}
\label{fig1}
\end{figure}

\section{Method}

\subsubsection{Problem Setup.} Given an fMRI pre-trained model $f(\cdot)$ with parameters $\theta$, and a high-resource fMRI dataset $\mathcal{D}=\{(\mathbf{x}_i,\mathbf{y}_i)\}$, where each data $\mathbf{x}_i$ is paired with $\mathbf{y}_i=[y_{(1,i)},y_{(2,i)},...,y_{(n,i)},...,y_{(N,i)}]$ corresponding to $N$ high-resource tasks $\mathcal{T}=\{T_1,...,T_n,...,T_N\}$, our goal is to learn a new low-resource task $T_{\text{target}}$ by efficiently updating parameters $\phi$ given the target task dataset $\mathcal{D}'=\{(\mathbf{x'},y')\}$ ($|\mathcal{D}|>|\mathcal{D}'|$). The number of updating parameters $\phi$ is much smaller than that of trainable parameters $\theta$ in the pre-trained model $f(\cdot)$.

\subsubsection{Overview.} Prompts in ScaPT are trainable embeddings that direct the model's responses without changing its architecture. Shown in Figure \ref{fig1}, ScaPT operates in two stages: Source Training (ST), in which trainable modular prompts (encoding abstract information and fundamental skills required for fMRI understanding) are interpolated to one participant-wise phenotype prompt. Participant-wise prompts corresponding to the same phenotype are averaged to formulate group-wise phenotype prompt, serving as knowledge base for the subsequent stage: Target Training (TT). It combines group-level phenotype prompts with a new target prompt to formulate a participant-wise vertex prompt for the target task. The weights for interpolation of prompts are from attention between projected input and prompts, after mapping the input to prompt space through Deeply-conditioned Input-Prompt (DIP) mapping.

\subsubsection{Deeply-conditioned Input-Prompt (DIP) Mapping.}

Previous research indicates that soft prompts might not match input embedding spaces \cite{khashabi2022prompt}, making direct attention between input and prompts unreliable. A proposed solution involves a network for projecting inputs into prompt spaces \cite{asai2022attempt}, but it falls short in multitasking scenarios by projecting inputs uniformly, thus failing to capture task-specific information. It also increases training parameters due to the addition of a separate network and relies on linear projections, which may not effectively represent the complex relationships between inputs and prompts.

To address the above issues, we propose Deeply-conditioned Input-Prompt (DIP) mapping $\mathcal{M}$ through reusing $f(\cdot)$ (Figure \ref{fig1}). We introduce learnable deep conditioning (DC) tokens $\mathcal{C}=\{\mathbf{c}_1,...,\mathbf{c}_n,\mathbf{c}_N|\mathbf{c}_n \in \mathbb{R}^{e\times m}\}$ that are prepended to the input, to guide $f(\cdot)$ to map the input to an appropriate prompt space conditioned on the given task. The prediction output from $f(\cdot)$ is then fed into a linear projection layer and a Layer Norm $\text{LN}(\cdot)$ to avoid gradient explosion \cite{asai2022attempt}. To deeply guide the conditional mapping from input to prompt spaces, we inject DC to every layer of $f(\cdot)$. Formally, the input-prompt mapping is defined as:

\begin{equation}
    \mathbf{H}_{(n,i)} = f([\mathbf{c}_n;\mathbf{X}_i]); \quad  
    \hat{\mathbf{H}}_{(n,i)} = \text{LN}(\text{NonLinear}(\mathbf{W}^{T}\mathbf{H}_{(n,i)})
\end{equation}

where $\mathbf{X}_i\in \mathbb{R}^{l\times m}$ is the ``text-like" representation generated from fMRI following \cite{thomas2022self}, describing the signal of each of $m$ brain networks for each time point $l$. $\mathbf{X}_i$ prepended by DC $[\mathbf{c}_{n};\mathbf{X}_i] \in \mathbb{R}^{(l+e)\times m}$ is input to the frozen $f(\cdot)$. $\mathbf{H}_{(n,i)} \in \mathbb{R}^{h}$ is the conditional output from $f(\cdot)$, with $h$ representing the hidden dimension. $\mathbf{W}\in \mathbb{R}^{h\times h}$ is the projection parameter to be updated during training,
and $\hat{\mathbf{H}}_{(n,i)} \in \mathbb{R}^{h}$ is the projected input.

\subsubsection{Source Training (ST).} A healthy cohort encompasses a large number of participants, with various phenotypes associated (high-resource tasks). In the first stage - Source Training (ST), we aim to train a set of phenotype prompts (PheP) that encapsulate information on different behavior-relevant brain phenotypes. It will serve as a source of knowledge for downstream tasks.

To better capture the relationship between input fMRI and prompts, and boost the capacity of PheP to match the complexity of input, we model each participant-wise PheP as an interpolation of a set of modular prompts (MoP), with the computed attention between input and MoP as the weights for interpolation, as shown in Figure \ref{fig1}. Each MoP can be seen as a basic skill; solving a task involves combining these fundamental skills \cite{sun2023multitask}. Formally, for a high-resource task $T_n \in \mathcal{T}$, our training objective is to maximize the likelihood of predicting the label $y_{(n,i)}$ as follows:

\begin{equation}
    \max_{\mathbf{P}_{(n,i)},\mathcal{M}} p_{\theta}(y_{(n,i)}|[\mathbf{P}_{(n,i)};\mathbf{X}_i])
\end{equation}

where $\mathbf{P}_{(n,i)} \in \mathbb{R}^{d\times m}$ is the participant-wise PheP with length $d$ for task $T_n$. $\mathbf{X}_i$ prepended by the prompt $[\mathbf{P}_{(n,i)};\mathbf{X}_i] \in \mathbb{R}^{(l+d)\times m}$ is input to the frozen $f(\cdot)$ to predict $\hat{y}_{(n,i)}$. $\mathbf{P}_{(n,i)}$ is generated from the interpolation of $K$ MoP:

\begin{equation}
    \mathbf{P}_{(n,i)}=\sum^{K}_{k=1}\alpha^{(n,i)}_k \cdot \mathbf{\dot{P}}_k, \, \alpha^{(n,i)}_k=\frac{e^{<\hat{\mathbf{P}}_k,\hat{\mathbf{H}}_{(n,i)}>/\tau}}{\sum_{i=1}^{K}e^{<\hat{\mathbf{P}}_i,\hat{\mathbf{H}}_{(n,i)}>/\tau}}; \quad \mathbf{\overline{P}}_n=\frac{1}{|\mathcal{D}|}\sum^{|\mathcal{D}|}_{i=1}\mathbf{P}_{(n,i)}
\end{equation}

where we compute the attention score $\alpha^{(n,i)}_k$ between MoP $\hat{\mathbf{P}}_k \in \mathbb{R}^{h}$ (max-pool of $\dot{\mathbf{P}}_k$ from $\mathbb{R}^{d\times m}$ to $\mathbb{R}^{m}$, followed by a linear transformation to $\mathbb{R}^{h}$) and $\hat{\mathbf{H}}_{(n,i)}$. $\tau$ is the temperature. $\mathbf{P}_{(n,i)}$ is a participant-wise prompt for $\mathbf{X}_i$. To transfer the learned prompts to the next stage, we average $\mathbf{P}_{(n,i)}$ to formulate one group-wise PheP $\mathbf{\overline{P}}_n$, which is utilized subsequently for next stage.

\subsubsection{Target Training (TT).} Prompts for a new task could be blended with pre-trained prompts to incorporate gained knowledge \cite{asai2022attempt}. In the second stage - Target Training (TT), we first initialize a target prompt $\mathbf{P}_{\text{target}}$ tailored for a target task. To capitalize on the insights embedded in $\mathbf{\overline{P}}_n$, we learn a vertex prompt $\mathbf{P}^*$, by interpolating $\mathbf{\overline{P}}_n$ and $\mathbf{P}_{\text{target}}$ given attention computed by $\mathcal{M}$ (Figure \ref{fig1}). Similar to ST, the goal of TT is to maximize the likelihood of predicting the correct target task label $y'$, given the concatenation of $\mathbf{P}^*$ and input $\mathbf{X'}$:

\begin{equation}
    \max_{\mathbf{P}_{\text{target}},\mathcal{M}} p_{\theta}(y'|[\mathbf{P}^*;\mathbf{X'}]); \quad \mathbf{P}^*=\mathbf{P}_{\text{target}}+\sum^{N}_{n=1}\beta_n \cdot \mathbf{\overline{P}}_n
\end{equation}

where $\mathbf{P}^*$ is the interpolation of $\mathbf{P}_{\text{target}}$ and $\mathbf{\overline{P}}_n$. $\beta_n$ is the attention score between $\mathbf{X'}$ and the $\mathbf{\overline{P}}_n$ computed by $\mathcal{M}$.

\section{Experiments}

\subsubsection{Datasets.} In ST, resting state fMRI data from 656 participants of the Lifespan Human Connectome Project Aging (HCP-A) \cite{bookheimer2019lifespan,harms2018extending} were analyzed to predict 38 phenotypes (see details in the supplementary material), alongside sex and age. The brain-behavior phenotypes established at this stage include three domains: cognition, personality, and social emotion. We hypothesized that ScaPT would perform well in tasks relevant to these domains. In TT, ScaPT was assessed on two classification tasks for neurodegenerative disease diagnosis/prognosis (related to cognition) using Alzheimer’s Disease Neuroimaging Initiative (ADNI) \cite{jack2008alzheimer}, and a regression task for personality trait prediction (related to personality and social emotion) using UK Biobank (UKB) \cite{sudlow2015uk}. FMRI in MNI space were preprocessed using fMRIPrep \cite{esteban2019fmriprep} (20.2.3) with the default settings and parcellated to 1024 networks using DiFuMo with  \href{https://nilearn.github.io/dev/modules/description/difumo_atlases.html}{Nilearn library} (0.10.4).

The experiments of TT were performed similar to \cite{thomas2022self}, using limited data to showcase the adaptation performance and varying the size to demonstrate how performance scales. (1) Control Normal (CN) v.s. Mild Cognition Impairment (MCI) classification: 340 ADNI participants, with 170 for each class. We first allocated an independent test set consisting of 100 CNs and 100 MCIs. From the remaining 140, we randomly sampled 3/5/10 data per class for training, using the rest for validation. Sampling was repeated 10 times, and accuracy and F1 score were reported with mean and standard deviation. (2) Amyloid Positive v.s. Negative classification: 100 ADNI participants with normal cognition (50 for each class). It was performed under the similar experimental setting with an independent test set consisting of 25 for each class. (3) Neuroticism Score Prediction: Neuroticism describes a tendency to experience unsettling feelings. The score ranges from 0 to 12 (normalized to 0-1 for training), with higher score corresponding to a greater tendency. We have 1000 UKB participants with neuroticism scores (800 for training and 200 for testing). Training samples were increased to 30/50/100 (randomly sampled from 800), following the same settings as the prior tasks.

\subsubsection{Training Details.} We adopt the state-of-the-art fMRI language model \cite{thomas2022self}, with causal sequence modeling structure, for our downstream adaptation. It was pre-trained using 11,980 runs of 1,726 individuals across 34 datasets. The pre-trained model contains 4 GPT-2 layers \cite{radford2019language}, with 12 attention heads in each self-attention module. To utilize the pre-trained model, the input must be parcellated by Dictionaries of Functional Modes (DiFuMo) \cite{dadi2020fine}. The preprocessed input to the model is $\mathbf{X}\in \mathbb{R}^{l\times m}$ obtained by DiFuMo, where $l$ is the input sequence length and $m=1024$ networks. The hidden dimension in the model is $h=768$. We utilized $K=5$ modular prompts. Prompt/DC length is $d=e=5$. Refer to supplementary for more hyper-parameters and experimental settings.


\subsubsection{Main Results.} In a scenario with limited training data, we evaluated ScaPT against fine-tuning and three multitask-based prompt-tuning approaches: SPoT \cite{vu2022spot}, MP² \cite{sun2023multitask}, and ATTEMPT \cite{asai2022attempt} (Table \ref{tab1} and \ref{tab2}). For fine-tuning, $f(\cdot)$ underwent direct fine-tuning using TT datasets (ADNI/UKB). Meanwhile, for the multitask-based methods, prompts were initially trained on ST datasets (HCP-A), which were then served as the prompt initialization in the TT stage. 

ScaPT demonstrated superior performance over both fine-tuning and other prompt tuning methods across various sizes of training datasets, scaling well with the number of training data. This underscores ScaPT's effectiveness in transferring knowledge from high-resource tasks to those with scarce resources.

{
\setlength{\tabcolsep}{2pt} 

\begin{table}[!t]
\centering
\caption{Accuracy (Acc) and F1 score on CN v.s. MCI and Amyloid $a\beta+ve$ v.s. $a\beta-ve$ classification (\%, mean(standard deviation) for 10 independent runs), trained on varying training dataset size ($I=3,5,10$ per class). The best results are in \textbf{bold}, with \textbf{*} denoting significant improvement ($p<0.05$).}
\label{tab:method_comparison}
\begin{tabular}{lcccccc}
\toprule
\multirow{2.5}{*}{Methods} & \multicolumn{2}{c}{$I=3$} & \multicolumn{2}{c}{$I=5$} & \multicolumn{2}{c}{$I=10$} \\
\cmidrule(lr){2-3} \cmidrule(lr){4-5} \cmidrule(lr){6-7}
& Acc $\uparrow$ & F1 $\uparrow$ & Acc $\uparrow$ & F1 $\uparrow$ & Acc $\uparrow$ & F1 $\uparrow$ \\
\midrule
\multicolumn{7}{l}{\emph{CN v.s. MCI}} \\
\midrule
Fine-tuning \cite{thomas2022self} & 53.0(3.1) & 47.3(3.7) & 61.1(3.9) & 57.1(5.4) & 65.3(2.7) & 62.0(3.2) \\
SPoT \cite{vu2022spot} & 56.3(.70) & 51.4(1.6) & 65.6(.35) & 62.9(.66) & 64.8(2.0) & 61.2(2.5) \\
$\text{MP}^2$ \cite{sun2023multitask} & 65.0(2.9) & 59.2(3.8) & 71.4(3.1) & 68.3(3.4) & 74.4(2.0) & 71.3(2.5) \\
ATTEMPT \cite{asai2022attempt} & 64.3(1.5) & 64.4(2.1) & 75.0(1.9) & \textbf{74.9}(2.0) & 77.3(2.0) & 76.5(2.2) \\
\rowcolor{lightgray}\textbf{ScaPT} & \textbf{69.7*}(2.8) & \textbf{67.4*}(2.9) & \textbf{75.1}(2.1) & 73.3(2.8) & \textbf{80.3*}(2.2) & \textbf{79.4*}(2.2) \\
\midrule
\multicolumn{7}{l}{\emph{Amyloid $a\beta+$ve v.s. $a\beta-$ve}} \\
\midrule
Fine-tuning \cite{thomas2022self} & 51.1(.15) & 52.3(1.1) & 64.2(0.2) & 66.1(3.2) & 71.1(.15) & 72.2(.01) \\
SPoT \cite{vu2022spot} & 55.9(.49) & 50.7(.88) & 58.9(.20) & 51.1(.71) & 64.5(.35) & 58.1(.26) \\
$\text{MP}^2$ \cite{sun2023multitask} & 51.3(2.2) & 53.2(1.2) & 70.6(.13) & 71.2(2.1) & 78.4(.15) & 79.1(.21) \\
ATTEMPT \cite{asai2022attempt} & 57.6(2.6) & 61.7(1.3) & 68.1(.37) & 70.2(.69) & 80.6(1.6) & 81.5(1.4) \\
\rowcolor{lightgray}\textbf{ScaPT} & \textbf{61.2*}(1.2) & \textbf{63.1*}(1.1) & \textbf{71.2}(.53) & \textbf{72.1*}(.25) & \textbf{86.0*}(.10) & \textbf{87.7*}(.10) \\
\bottomrule
\end{tabular}
\label{tab1}
\end{table}

} 

{
\setlength{\tabcolsep}{1.2pt} 

\begin{table}[!h]
\centering
\caption{Mean Absolute Error (MAE,$\times 10^{-1}$) and Pearson Correlation ($\rho$) on Neuroticism score prediction.}
\label{tab:method_comparison}
\begin{tabular}{lcccccc}
\toprule
\multirow{2.5}{*}{Methods} & \multicolumn{2}{c}{$I=30$} & \multicolumn{2}{c}{$I=50$} & \multicolumn{2}{c}{$I=100$} \\
\cmidrule(lr){2-3} \cmidrule(lr){4-5} \cmidrule(lr){6-7}
& MAE $\downarrow$ & $\rho$ $\uparrow$ & MAE $\downarrow$ & $\rho$ $\uparrow$ & MAE $\downarrow$ & $\rho$ $\uparrow$ \\
\midrule
Fine-tuning \cite{thomas2022self} & 3.03(.43) & 0.37(.008) & 2.57(.21) & 0.40(.002) & 2.35(.00) & 0.45(.002) \\
SPoT \cite{vu2022spot} & 5.29(.74) & 0.35(.005) & 4.43(.69) & 0.38(.004) & 4.37(.58) & 0.39(.003) \\
$\text{MP}^2$ \cite{sun2023multitask} & 3.92(.15) & \textbf{0.40}(.004) & 3.59(.16) & 0.41(.003) & 3.41(.12) & 0.42(.002) \\
ATTEMPT \cite{asai2022attempt} & 2.85(.18) & 0.38(.004) & \textbf{2.37}(.19) & 0.43(.006) & 2.11(.21) & 0.45(.004) \\
\rowcolor{lightgray}\textbf{ScaPT} & \textbf{2.53*}(.27) & \textbf{0.40}(.003) & 2.40(.25) & \textbf{0.45*}(.001) & \textbf{1.90*}(.06) & \textbf{0.49*}(.003) \\
\bottomrule
\end{tabular}
\label{tab2}
\end{table}

} 

\begin{figure}[t]
\centering
\includegraphics[width=\linewidth]{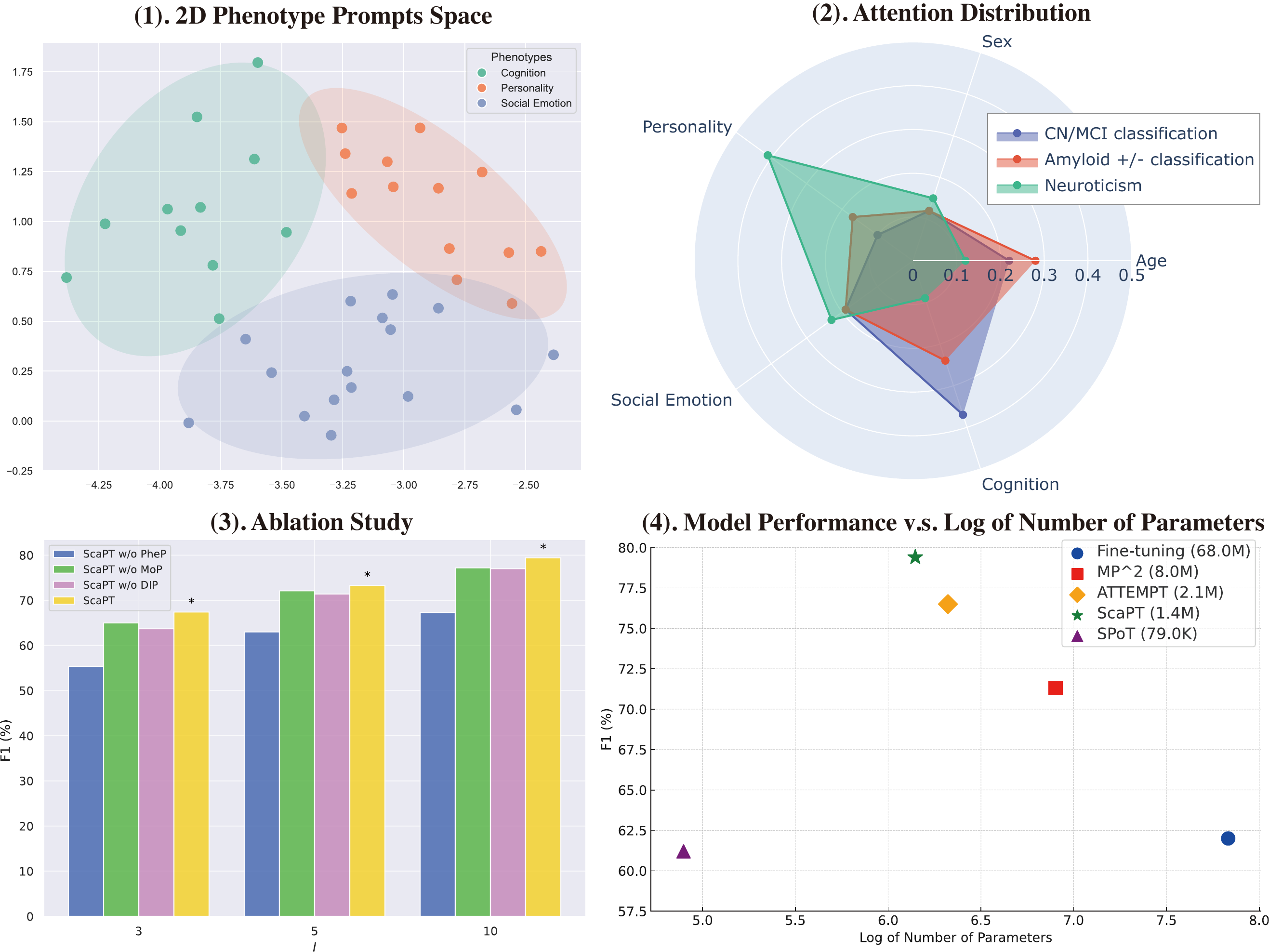}
\caption{Further analysis: (1) visualization of 2D phenotype prompts space, (2) interpretation of target tasks through attention distribution, (3) ablation study, and (4) comparison of model performance versus numbers of parameters.}
\label{fig2}
\end{figure}

\subsubsection{Prompt Interpretation.} In the ST stage, we created 40 phenotype prompts for 38 phenotype predictions, alongside age and sex determinations. Remarkably, these prompts naturally formed into three clusters - Personality, Social Emotion, and Cognition - without prior supervision, indicating they effectively capture different pillars of brain-behavior associations (Figure \ref{fig2}-1). 

During the TT stage, the attention scores between input and phenotype prompts (Figure \ref{fig2}-2) aid in interpreting the target task. Attention score vectors were averaged across inputs, and then attributes within each group were averaged (without the one for $\mathbf{P}_{\text{target}}$). These five values were normalized for analysis. Aligning well with the literature, ScaPT shows a focus on ``cognition" or ``age" in neurodegenerative disease diagnosis/prognosis task, while it focuses on ``personality" for neuroticism score prediction.

\subsubsection{Ablation Study and Parameter-efficiency.} We evaluated ScaPT against its ablations (Figure \ref{fig2}-3), including ScaPT w/o PheP (using MoP directly for $\mathbf{P^*}$ formulation without high-resource task training), ScaPT w/o MoP (learning $\mathbf{\overline{P}}_n$ without prompt width expansion), and ScaPT w/o DIP (utilizing ATTEMPT's subnetwork for input-prompt mapping). The absence of PheP led to a significant performance drop, underscoring the importance of high-resource task knowledge in boosting low-resource task performance. ScaPT outperformed its counterparts lacking MoP, demonstrating MoP's role in enhancing expressive capacity by widening prompts. Additionally, ScaPT's DIP module surpassed ATTEMPT's subnetwork in mapping inputs to prompts, effectively capturing complex input-prompt relationships using $f(\cdot)$. In Figure \ref{fig2}-4, we compared ScaPT's performance with other models relative to their trainable parameters. ScaPT significantly outperformed fine-tuning, $\text{MP}^2$, and ATTEMPT, despite updating only 2\% of total parameters. Although SPoT had the fewest trainable parameters, its performance lagged, likely due to its limited feature capture capability.

\section{Conclusion}

We introduce Scaffold Prompt Tuning (ScaPT), the first prompt-based adaptation framework for large-scale fMRI pre-trained models, compatible with very limited training data. ScaPT features a hierarchical prompt structure that facilitates knowledge transfer from high-resource tasks to those with fewer resources. Moreover, we develop a Deeply-conditioned Input-Prompt (DIP) mapping to capture the complex relationship between the input and prompt spaces. Our experiments demonstrate ScaPT's exceptional parameter efficiency and its superior performance in neurodegenerative disease diagnosis or prognosis, as well as personality trait prediction from resting-state fMRI data. In addition, our attention mechanism offers semantic interpretation for target tasks. Future studies could expand ScaPT's reach to longitudinal data, and explore its possibility for prompting vision models.

\begin{credits}
\subsubsection{\ackname} This study was supported by the Singapore National Medical Research Council (NMRC\textbackslash{}OFLCG19May-0035, NMRC\textbackslash{}CIRG\textbackslash{}1485\textbackslash{}2018, NMRC\textbackslash{}CSA-SI\textbackslash{}0007\textbackslash{}2016, NMRC\textbackslash{}MOH-00707-01, NMRC\textbackslash{}CG\textbackslash{}435 M009\textbackslash{}2017-NUH\textbackslash{}NUHS, \\CIRG21nov-0007 and HLCA23Feb-0004), RIE2020 AME Programmatic Fund from \\A*STAR (No. A20G8b0102), Ministry of Education (MOE-T2EP40120-0007 \& T2EP2-0223-0025, MOE-T2EP20220-0001), and Yong Loo Lin School of Medicine Research Core Funding, National University of Singapore, Singapore. Yueming Jin was supported by MoE Tier 1 Start up grant (WBS: A-8001267-00-00).

\subsubsection{\discintname}
The authors have no competing interests to declare that are
relevant to the content of this article.
\end{credits}

%
%
%
\bibliographystyle{splncs04}
\bibliography{ref}

\end{document}